\documentclass[12pt,a4paper]{article}
\usepackage{feynmp,amsmath,amssymb}
\usepackage{array}
\usepackage{thophys}
\makeatletter
\newif\if@preliminary
\@preliminaryfalse
\def\preliminary{\@preliminarytrue}
\def\preprintno#1{\def\@preprintno{#1}}
\def\address#1{\def\@address{#1}}
\def\email#1#2{\thanks{\tt #1{}#2}}
\def\abstract#1{\def\@abstract{#1}}
\renewcommand\abstractname{ABSTRACT}
\newlength\preprintnoskip
\newlength\abstractwidth
\@titlepagetrue
\renewcommand\maketitle{\begin{titlepage}%
  \let\footnotesize\small
  \hfill\parbox{\preprintnoskip}{%
  \begin{flushright}\@preprintno\end{flushright}}\hspace*{1cm}
  \vskip 60\p@
  \begin{center}%
    {\Large\bf\boldmath \@title \par}\vskip 10mm%
    {\sc\@author \par}\vskip 8mm%
    {\@address \par}%
    \if@preliminary
      \vskip 2cm {\large\sf PRELIMINARY DRAFT \par \@date}%
    \fi
  \end{center}\par
  \@thanks
  \vfill
  \begin{center}%
    \parbox{\abstractwidth}{\centerline{\abstractname}%
    \vskip 3mm%
    \@abstract}
  \end{center}
  \end{titlepage}%
  \setcounter{footnote}{0}%
  \let\thanks\relax\let\maketitle\relax
  \gdef\@thanks{}\gdef\@author{}\gdef\@address{}%
  \gdef\@title{}\gdef\@abstract{}\gdef\@preprintno{}
}%
\def\@citex[#1]#2{\if@filesw\immediate\write\@auxout{\string\citation{#2}}\fi
  \def\@citea{}\@cite{\@for\@citeb:=#2\do
    {\@citea\def\@citea{,\penalty\@m}\@ifundefined
       {b@\@citeb}{{\bf ?}\@warning
       {Citation `\@citeb' on page \thepage \space undefined}}%
\hbox{\csname b@\@citeb\endcsname}}}{#1}}
\def\citerange{\@ifnextchar [{\@tempswatrue\@citexr}{\@tempswafalse\@citexr[]}}
\def\@citexr[#1]#2{\if@filesw\immediate\write\@auxout{\string\citation{#2}}\fi
  \def\@citea{}\@cite{\@for\@citeb:=#2\do
    {\@citea\def\@citea{--\penalty\@m}\@ifundefined
       {b@\@citeb}{{\bf ?}\@warning
       {Citation `\@citeb' on page \thepage \space undefined}}%
\hbox{\csname b@\@citeb\endcsname}}}{#1}}
\long\def\@makecaption#1#2{%
  \vskip\abovecaptionskip
  \sbox\@tempboxa{#1: \emph{#2}}%
  \ifdim \wd\@tempboxa >\hsize
    #1: \emph{#2}\par
  \else
    \hbox to\hsize{\hfil\box\@tempboxa\hfil}%
  \fi
  \vskip\belowcaptionskip}
\def\fmslash{\@ifnextchar[{\fmsl@sh}{\fmsl@sh[0mu]}}
\def\fmsl@sh[#1]#2{%
  \mathchoice
    {\@fmsl@sh\displaystyle{#1}{#2}}%
    {\@fmsl@sh\textstyle{#1}{#2}}%
    {\@fmsl@sh\scriptstyle{#1}{#2}}%
    {\@fmsl@sh\scriptscriptstyle{#1}{#2}}}
\def\@fmsl@sh#1#2#3{\m@th\ooalign{$\hfil#1\mkern#2/\hfil$\crcr$#1#3$}}
\makeatother

\newcommand{\Op}{\mathcal{O}}

\newcommand{\ii}{\mathrm{i}}

\newcommand{\ap}{\frac{\alpha_s}{\pi}}
\newcommand{\tg}{\tilde{\gamma}}
\newcommand{\bq}{\overline{B}{}^0}
\renewcommand{\Re}{\text{Re}}
\renewcommand{\Im}{\text{Im}}
\begin{document}
\preprintno{SI-HEP-2004-04\\TTP03--23\\
            hep-ph/0403085\\
  [0.5\baselineskip] March 2004}
\title{%
The Gold-plated mode revisited: \\
$\sin(2\beta)$ and $B^0 \to J/\Psi\,K_S$ in the Standard Model
}
\author{%
 Heike~Boos$^\dag$\email{boos}{@particle.uni-karlsruhe.de},
            Thomas~Mannel$^*$\email{mannel}{@hep.physik.uni-siegen.de}, 
 and J\"urgen~Reuter$^\dag$\email{reuter}{@particle.uni-karlsruhe.de}
}
\address{\it
 $^*$ Theoretische Physik 1, Fachbereich Physik, University of Siegen, \\
 D--57068 Siegen, Germany \\[2mm]
 $^\dag$ Institut f\"ur Theoretische Teilchenphysik, University of Karlsruhe,\\
 D--76128 Karlsruhe, Germany
}
\abstract{%
  We study the corrections to the determination of $\sin(2\beta)$ from
  the time dependent CP asymmetry of $B^0 \to J/\Psi\,K_S$ which arise in
  the standard model. Although a precise prediction of these corrections
  is not possible we find that they are indeed extremely small, of the
  order of less than a per mil of the observed value. This 
  means in turn that any deviation
  visible at the $B$ factories will be a clear signal for new physics.} 
\maketitle

\begin{fmffile}{\jobname pics}
\section{Introduction}
The measurement of the mixing-induced CP asymmetry in the so-called
gold-plated mode $B^0 \to J/\Psi\,K_S$ is becoming a precision
measurement \cite{sin2beta}. Currently the relative uncertainty is 
at the level of five
percent and is going to decrease further in the next few years
as more data from the $B$ factories are analyzed. 

From the theoretical side the time dependent CP asymmetry in this channel
is related in the standard model to $\sin(2\beta)$ in a very clean way \cite{BigiSanda},
i.e.~it is not plagued by
hadronic uncertainties, at least at the level of current experimental
precision. However, the precision of the experimental data will increase
further, making this CP asymmetry an interesting probe for new physics.  

Possible new physics effects have been discussed in some detail in a generic
framework in \cite{FleischerMannel} where both the charged and the neutral
modes of $B \to  J/\Psi\,K $ have to be taken into account. In
\cite{FleischerMannel} certain observables have been defined which are
sensitive to different aspects of new physics. 

However, in order to quantitatively pin down a possible new physics
effect and to assess the reach to new physics, the small
standard-model contributions have to be under control. Although 
estimates of these effects have been given some time ago \cite{oldstuff},
it is worthwhile to reconsider these estimates motivated by the experimental 
precision to be expected soon. 

In this paper we shall 
try to get a quantitative estimate for
the time-dependent CP asymmetry in this channel beyond the
simple relations (in the convention used in  (\ref{cpasymm0}) below)
$$
C_{B\to J/\Psi\,K_S} = 0 \qquad
S_{B\to J/\Psi\,K_S} = \sin(2\beta) \, . 
$$
The corrections to these relations originate from two sources. The first source
is from the corrections to the ${\Delta B=\pm 2}$ part of the Hamiltonian. 
In the present paper we compute these contributions
systematically in an effective field theory by
subsequently integrating out heavy degrees of freedom down to a small 
hadronic scale. It is worthwhile to note that many elements of our calculation 
are similar to the calculation of the lifetime difference in the  system of 
neutral $B$ mesons which are performed in an effective field theory framework in 
\cite{Life1}.  

The second source is the decay amplitude itself which has
small contributions carrying a different CP phase compared to the leading
piece. The latter is much harder to estimate and we shall refer to
well known methods. 

A full discussion of the standard-model effects would also require to include 
the effects from CP violation in the Kaon system. We shall not discuss these
effects in the present paper, we focus completely on the effects coming from the 
$B$ system 

In the next sections we set up the calculation of these two contributions
and discuss the methods of calculation. Finally we summarize our results
and conclude.

\section{Basic relations} 

CP asymmetries are measured at the $B$ factories by detecting
the decay products of the coherent $B^0 \bq$ pair. One of them is
identified as a flavor state using e.g.~a leptonic tagging mode while
the decay of the other into a CP eigenstate is observed. Usually, in
the calculation of these rates a possible CP violation stemming from
the mixing on the tagging side or in the tagging decay is neglected 
which results in the well-known formula for the decay 
$B^0 \to J/\Psi K_S$
\begin{equation}
  \label{cpasymm0}
  \left[a^{B\to J/\Psi\,K_S}_{CP}\right]_0 (t) = \frac{
  C_{B\to J/\Psi\,K_S}  \cos(\Delta m \, t)
  - S_{B\to J/\Psi\,K_S} \sin(\Delta m \, t)}
    {\cosh(\Delta\Gamma \,t/2)
         + D_{B\to J/\Psi\,K_S} \sinh (\Delta\Gamma \, t/2)} 
\end{equation}
with
\begin{equation}
  C_{B\to J/\Psi\,K_S} = \frac{1 -
  |\lambda|^2}{1 + |\lambda|^2}, \quad
  S_{B\to J/\Psi\,K_S} = \frac{2 \, \Im [\lambda]}{1 + |\lambda|^2} , \quad 
  D_{B\to J/\Psi\,K_S} = \frac{2 \,  \Re [\lambda]}{1 + |\lambda|^2} 
\end{equation}
and
\begin{equation}
  \lambda = \left( \frac{q}{p} \right)_B \cdot \left( \frac{p}{q}
  \right)_K \cdot \frac{A(\bq \to J/\Psi K_S)}{A(B^0 \to J/\Psi
  K_S)} .
\end{equation}
Here $p$ and $q$ are the parameters characterizing the mixing in the $K$- and
$B$-system, while $A(B^0 \to J/\Psi K_S)$ is the amplitude for the decay
into the CP eigenstate $J/\Psi K_S$. 

Our aim is to investigate all effects correcting the leading
contribution and therefore we take also into account the influence of
the mixing 
in the $B$ system on the tagging amplitude assuming that there is no CP
violation in the tagging decay (cf.~\cite{tagcp}). Defining 
\begin{equation}
  \label{defepsilon}
  |\lambda_{\text{tag}}|^2 \equiv \left| \left( \frac{q}{p} \right)_B
  \cdot \frac{\overline{A}_{\text{tag}}}{A_{\text{tag}}} \right|^2 \approx
  \left|  \left( \frac{q}{p} \right)_B \right|^2 =: 1 + \epsilon,
\end{equation}
the CP asymmetry becomes 
\begin{equation}
  \label{cpasymmfull}
  a^{B\to J/\Psi\,K_S}_{CP}(t) = \left[a^{B\to
  J/\Psi\,K_S}_{CP}\right]_0(t) +\frac{\epsilon}2 - \frac{\epsilon}2  
  \sin^2(\Delta m \, t)\sin^2(2\beta).
\end{equation}

These relations above are true in general and take into account a possible
width difference, a possible direct CP asymmetry as well as the
mixing effects on the tagging. The lifetime difference $\Delta \Gamma$
as well as $1-|\lambda|^2$ and $\epsilon$ are small and the leading
term is obtained by neglecting these quantities. 
Furthermore, since the weak phase of the
decay amplitude of the leading contribution vanishes in the standard
convention, $S$ measures the weak phase of the $\Delta B = 2$ 
contribution to the effective Hamiltonian which is -- again to leading order 
-- simply $2 \beta$.

We make use of the general relation
\begin{equation}
  \left(C_{B\to J/\Psi\,K_S} \right)^2 +
  \left(S_{B\to J/\Psi\,K_S} \right)^2 + 
  \left(D_{B\to J/\Psi\,K_S} \right)^2 = 1,
\end{equation}
which allows us to replace
$D_{B\to J/\Psi\,K_S}$ by its leading order
expression $D_{B\to J/\Psi\,K_S} = \cos(2\beta)$,
since it is multiplied by the small
quantity $\Delta \Gamma \, t$ in the expansion. Note that typically
$t$ is of the order of the $B$ meson lifetime $\tau$ and we have
$\Delta \Gamma \, \tau \ll 1$.


\section{Corrections to the mixing}
The phenomenon of mixing between $B$ and $\overline{B}$ is due to the
box diagrams with a double $W$ exchange in the full electroweak
theory. These diagrams have been evaluated some time ago in the full standard
model including the quark masses of all quarks in the loop
\cite{Buras_und_co}. After GIM cancellation, the leading term is due to the
top quark, giving rise to a contribution of the order $(m_t / M_W)^2$ with the
weak phase $2 \beta$. Subleading terms are either of the order $(m_b / M_W)^2$
which again carry the phase $2 \beta$ and thus will not contribute to a
modification of $S$, or of order $(m_c / M_W)^2$
which carry a different weak phase and hence will yield a correction to
$S$.

Instead of calculating the box diagrams in the full theory one may
also use an effective theory picture which we will do in this paper.
The approach is similar to the one used for $D-\overline{D}$ mixing
\cite{Bigiddbar,ddbar,ddbar1}, and also for the $\Delta S = 2$ Hamiltonian
in \cite{NiersteHerrlich}. 
The advantage is that one may express 
the contributions to the mixing phase in terms of matrix elements of
certain local operators, which can be estimated in factorization. Furthermore, 
one can resum systematically large logarithms using the renormalization 
group. 

The first step is the usual one, in which the $W$ boson and the top quark
are integrated out at a common scale $\mu \sim m_t \sim M_W$. The result
is the well known $\Delta B = 2$ operator of dimension six, 
which is
\begin{equation} \label{DB2}
H^{\Delta B=2}_{\text{eff}} = \frac{G^2_F}{4 \pi^2}
\lambda_t^2 m_t^2 C_0(x_t) Q_0  \qquad \mbox{with} \qquad 
Q_0 = (\bar{b}_L \gamma_\mu d_L) (\bar{b}_L \gamma^\mu d_L)
\end{equation}
where $\lambda_q = V_{qb}^* V_{qd}$, $x_t = m_t^2 / M_W^2$ and
\begin{equation}
  C_0 (x_t) = \frac{4-11 x_t + x_t^2}{4(1-x_t)^2}
            - \frac{3 x_t^2 \ln x_t}{2(1-x_t)^3}
  \end{equation}
is obtained from the usual Inami-Lim function \cite{InamiLim}.
Note that we extracted explicitly a factor $m_t^2$ which is a
reminiscent of the GIM mechanism. 

However, (\ref{DB2}) is not the only contribution to $\Delta B = 2$. Other
contributions originate from two insertions of $\Delta B = 1$
operators 
\begin{equation}
T^{\Delta B=2} = -\frac{i}{2} \int d^4 x \,
T \left[H^{\Delta B=1}_{\text{eff}} (x) H^{\Delta B=1}_{\text{eff}}
  (0) \right],
\end{equation}
where the relevant operators are
\begin{eqnarray}
H^{\Delta B=1}_{\text{eff}} = \frac{4 G_F}{\sqrt{2}} && \!\!\!\!\!\!\!
\left[V_{cb}^* V_{cd} (\bar{b}_L \gamma_\mu c_L) (\bar{c}_L \gamma^\mu d_L) 
     +   V_{cb}^* V_{ud} (\bar{b}_L \gamma_\mu c_L) (\bar{u}_L \gamma^\mu d_L)
     \right.  \\ \nonumber  
&&   \left.
     +   V_{ub}^* V_{cd} (\bar{b}_L \gamma_\mu u_L) (\bar{c}_L \gamma^\mu d_L)
     +   V_{ub}^* V_{ud} (\bar{b}_L \gamma_\mu u_L) (\bar{u}_L \gamma^\mu d_L)
	 \right]
\end{eqnarray} 
which yield non-local contributions at the scale $\mu \sim M_W$.

Furthermore, the matching at the scale $M_W$ yields -- aside from the above
dimension-6 operator appearing in (\ref{DB2}) -- dimension-8
contributions involving two covariant derivatives. These 
have been partially calculated in \cite{KraussSoff}; however, we do not
need these contributions for our purposes, since these pieces are again
dominated by the top quark and thus have the same weak phase as
the leading part.

In addition, keeping a non-zero charm mass, the matching calculation yields
an operator of the form
$m_c^2 (\bar{b}_L \gamma_\mu d_L) (\bar{b}_L \gamma^\mu d_L)$
which we shall treat as a dimension-8 operator as well. The matching of this
operator involves the calculation of the box diagrams keeping the charm mass
non-zero. 

Lowering the scale turns the high momentum part of the non-local 
contributios into local operators. This effect is described by a
renormalization-group mixing of the non-local operators into local ones.
We define the non-local operators as 
\begin{subequations}
\begin{eqnarray}
T_1 &=&  -\frac{i}{2} \int d^4 x \,
T \left[ (\bar{b}_L \gamma_\mu c_L )(x) (\bar{c}_L \gamma^\mu d_L )(x)
         (\bar{b}_L \gamma^\nu c_L )(0) (\bar{c}_L \gamma_\nu d_L )(0) \right]
\\
  T_2 &=&  -\frac{i}{2} \int d^4 x \,
T \left[ (\bar{b}_L \gamma_\mu c_L )(x) (\bar{u}_L \gamma^\mu d_L )(x)
         (\bar{b}_L \gamma^\nu u_L )(0) (\bar{c}_L \gamma_\nu d_L )(0) \right]
\\
  T_3 &=&  -\frac{i}{2} \int d^4 x \,
T \left[ (\bar{b}_L \gamma_\mu u_L )(x) (\bar{u}_L \gamma^\mu d_L )(x)
         (\bar{b}_L \gamma^\nu u_L )(0) (\bar{u}_L \gamma_\nu d_L )(0) \right]
 \end{eqnarray}
\end{subequations}
and find that these operators mix already at order $(\alpha_s)^0$ 
into local $\Delta B = 2$ operators of dimension~8\footnote{If the complete
       matching at $\mu = M_W$ would be performed, many more dim-8 operators
       would be induced; however, their coefficients would not be enhanced by
       a large logarithm; large logarithms at order $(\alpha_s)^0$ are only induced by
       the mixing with the non-local terms, which induces only $Q_1$ and
       $Q_2$.}
\begin{eqnarray}
 \label{q1q2q3}
  Q_1 &=& \Box (\bar{b}_L \gamma_\mu d_L )
           (\bar{b}_L \gamma^\mu d_L )\nonumber \\ 
  Q_2 &=& \partial^\mu \partial^\nu (\bar{b}_L \gamma_\mu d_L )
          (\bar{b}_L \gamma_\nu d_L ) \\ \nonumber
  Q_3 &=& m_c^2 (\bar{b}_L \gamma_\mu d_L ) (\bar{b}_L \gamma^\mu d_L ) .
  \end{eqnarray} 
This order $(\alpha_s)^0$ mixing happens via the diagrams
\vspace{1mm}
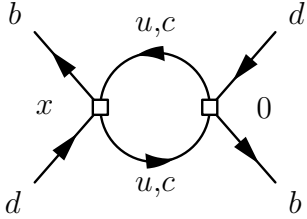
\begin{equation*}
\parbox{40\unitlength}{{}\hfil\\%
\begin{fmfgraph*}(40,20)
 \fmfleft{i1,i2}
      \fmfright{o1,o2}
 \fmflabel{$b$}{i2}
 \fmflabel{$b$}{o1}
 \fmflabel{$d$}{i1}
 \fmflabel{$d$}{o2}
 \fmflabel{$x\quad$}{v1}
 \fmflabel{$\quad 0$}{v2}
      \fmf{fermion}{i1,v1}
      \fmf{fermion}{v1,i2}
      \fmf{fermion}{v2,o1}
      \fmf{fermion}{o2,v2}
      \fmf{fermion,right,tension=0.6,label=$u$,,$c$}{v1,v2}
      \fmf{fermion,right,tension=0.6,label=$u$,,$c$}{v2,v1}
      \fmfv{decor.shape=square,decor.filled=empty,decor.size=2mm}{v1,v2}
\end{fmfgraph*}\\\hfil{}} \qquad\qquad .
\end{equation*}
The anomalous dimension matrix at order $(\alpha_s)^0$ is 
\begin{eqnarray}
\gamma = \frac{1}{48 \pi^2}
    \left(\begin{array}{cccccc}
          0 & 0 & 0 & 0 & 0 & 0 \\
	  0 & 0 & 0 & 0 & 0 & 0 \\
	  0 & 0 & 0 & 0 & 0 & 0 \\
          1 & 2 & 6 & 0 & 0 & 0 \\
          1 & 2 & 3 & 0 & 0 & 0 \\
	  1 & 2 & 0 & 0 & 0 & 0
        \end{array}\right),
\end{eqnarray}
and the operators are gathered into $\vec{O} =  (Q_1, Q_2, Q_3, T_1,
T_2, T_3)^T$.

The initial conditions for the renormalization-group running are from
matching. As discussed above, we do not explicitly need the matching
conditions for $Q_1$ and $Q_2$, so we get  
\begin{eqnarray}
  \hat{C}_1 (M_W, x_t) &=& -\frac{\lambda_t^2}{96\pi^2}C_{12} (x_t),        \qquad           
  \hat{C}_2 (M_W, x_t)  =  -\frac{\lambda_t^2}{48\pi^2}C_{12}^\prime
  (x_t), \nonumber  \\  
  \hat{C}_3 (M_W, x_t) &=& \frac{1}{32\pi^2} 
                           \left( 2 \lambda_c \lambda_t C_3 (x_t) + \lambda_{c}^2 \right), 
			    \\ \nonumber
  \hat{C}_4 (M_W) &=& \lambda_{c}^2,   \qquad \hat{C}_5 (M_W) = 2 \lambda_c \lambda_u, 
  \qquad \hat{C}_6 (M_W) = \lambda_u^2,
\end{eqnarray} 	      
where we have used the tree-level matching of the $\Delta B = 1$ operators
for the non-local terms. The function 
\begin{equation}
\ C_3 (x_t) = \ln x_t - \frac{3 x_t}{4(1-x_t)} - \frac{3 x_t^2 \ln x_t}{4(1-x_t)^2}
\end{equation}	      
is derived from the Inami-Lim function; $C_{12}$ and $C_{12}^\prime$
are functions of $x_t$ which could be obtained from the matching of
the dim-8 operators with two derivatives. We do not give these
functions here, since they are not needed in the following.  

The next step is to perform the renormalization-group running using the 
renormalization-group functions at order $(\alpha_s)^0$
\begin{equation}
\left( \mu \frac{\partial}{\partial \mu} - \gamma^T \right) \vec{C} (\mu) = 0
\end{equation}
where $\vec{C} = (\hat{C}_1, \hat{C}_2, \hat{C}_3, \hat{C}_4, \hat{C}_5, \hat{C}_6)$
	    are the coefficients of the operators.  

Running down to the scale of the $b$ quark mass we switch at $\mu \sim m_b$
again to another effective field theory in which the $b$ quark becomes static.
At this scale one has to replace the derivatives acting on the $b$ quark
field by \cite{HQET} 
\begin{equation}
  i \partial_\mu b \longrightarrow (m_b v_\mu + i \partial_\mu) h_v  
\end{equation}
where $h_v$ is the static $b$ quark field moving with velocity $v$.
Keeping only the leading term in the $1/m_b$ expansion we have to match onto
the operators
\begin{equation}
\begin{split}
  P_0 &=\;  (\bar{h}_{v,L}^{(+)} \gamma_\mu d_L )
                (\bar{h}_{v,L}^{(-)} \gamma^\mu d_L ) \\ 
  P_1 &=\; m_b^2 (\bar{h}_{v,L}^{(+)} \gamma_\mu d_L )
                (\bar{h}_{v,L}^{(-)} \gamma^\mu d_L ) \\ 
  P_2 &=\; m_b^2 (\bar{h}_{v,R}^{(+)} d_L )
                (\bar{h}_{v,R}^{(-)} d_L ) \\ 
  P_3 &=\; m_c^2 (\bar{h}_{v,L}^{(+)} \gamma_\mu d_L )
                (\bar{h}_{v,L}^{(-)} \gamma^\mu d_L ) ,
\end{split}
\end{equation}
where $h_{v,L}^{(+/-)}$ denotes the static quark/antiquark field,
which have become completely different fields in the static limit.

Performing the matching and the renormalization-group running we may evolve
down to scales around the charm-quark mass. At such a low scale all contributions
become local, once the up-quark mass is neglected. In fact the dim-8 operators
of the generic structure
$$
(\bar{h}_{v,R}^{(+)} D  d_L ) (\bar{h}_{v,R}^{(-)} D d_L ) \, ,
$$
with $D$ being a derivative inserted in any possible way, will yield
contributions proportional to $\Lambda_{\rm QCD}^2$. Compared to the pieces
considered here, these are suppressed by a factor $\Lambda_{\rm QCD}^2
/m_c^2 $ and thus may be neglected. 

Inserting back all the CKM factors, we make use of CKM unitarity, i.e.~the fact
that $\lambda_u + \lambda_c + \lambda_t = 0$. This separates the running
from CKM factors, which means that each weak phase is multiplied by a renormalization-group
invariant. Thus the $\Delta B = 2$ contributions  at the scale $\mu \sim m_c$
can be expressed in terms of the local effective Hamiltonian
\begin{multline} \label{simple}
  \mathcal{H}^{\Delta B=2}_{\text{eff}} = \frac{G^2_F}{4 \pi^2} \Biggl\{
   \lambda_t^2 m_t^2 C_0(x_t) P_0 
 + \frac{\lambda_t^2}{3} 
   \left[C_{12} (x_t)  - \ln\left( \frac{m_c^2}{M_W^2}\right) 
   \right] P_1  \\ - \frac{2\lambda_t^2}{3}  
   \left[C_{12}^\prime  (x_t)
         - \ln\left( \frac{m_c^2}{M_W^2}\right)
   \right] P_2   + \left(2\lambda_c \lambda_t 
   \left[C_3 (x_t)  - \ln\left( \frac{m_c^2}{M_W^2} \right)
   \right] + \lambda_c^2 \right) P_3 \Biggr\} .
\end{multline}
The first term in this expression is the well known leading term,
while two of the 
subleading terms are proportional to $m^2_b$ and carry the same weak
phase as the leading 
term. In these contributions we have the two unknown matching functions $C_{12}$ and
$C_{12}^\prime$; however, the running gives the large logarithm which in this framework
is assumed to dominate. Furthermore, the renormalization group reproduces the well known result for the term proportional to $m^2_c$, which will modify the
mixing phase.  Note that this effective Hamiltonian is given in terms of local 
operators $P_i$. We shall estimate the matrix elements of this
effective Hamiltonian using naive factorization at the scale $\mu \sim m_c$. 

From this effective Hamiltonian we can obtain the correction to the mixing
phase. Numerically we find 
\begin{multline}
  \label{simple2}
  \Delta \Im\left[\frac{M_{12}}{|M_{12}|} \right] = \frac{1}{C_0(x_t)}
  \frac{m_c^2}{m_t^2} \frac{|V_{cd}|
  |V_{cb}|}{|V_{td}| |V_{tb}|}\\
  \left[ \frac{|V_{cd}|
  |V_{cb}|}{|V_{td}| |V_{tb}|} \sin(2\beta)\cos(2\beta)
  + 2\left( C_3(x_t) -\ln\left(\frac{m_c^2}{M_W^2}\right)   \right)
  \left(\sin\beta - \sin(2\beta) \cos\beta\ \right)
   \right]  \\ 
  = -(4.48 \pm 2.55 )
  \cdot 10^{-4} .
\end{multline}

At tree level, the next order in the $1/m_b$ expansion consists of
operators 
of dimension-9, which are six-quark operators
(cf.~e.~g.~\cite{ddbar}). They originate from diagrams 
like ($\psi \equiv u,c$) 
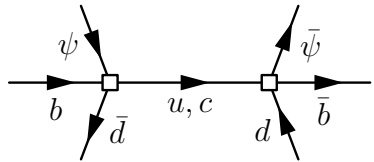
\begin{equation*}
  \parbox{55\unitlength}{%
      \begin{fmfgraph*}(48,22)
        \fmfleft{l}
        \fmfright{r}
	\fmftop{t1,t2,t3,t4,t5,t6}
	\fmfbottom{b1,b2,b3,b4,b5,b6}
	\fmf{fermion,label=$\bar{d}$,l.side=left}{v1,b2}
	\fmf{fermion,label=$b$}{l,v1}
	\fmf{fermion,label=$\psi$}{t2,v1}
	\fmf{fermion,label=$\bar{b}$}{v2,r}
	\fmf{fermion,label=$d$}{b5,v2}
	\fmf{fermion,label=$\bar{\psi}$}{v2,t5}
	\fmf{fermion,label=$u,,c$}{v1,v2}	
        \fmfv{decor.shape=square,decor.filled=empty,decor.size=2mm}{v1,v2}
      \end{fmfgraph*}} \qquad .
\end{equation*}
The contributions from these operators can be brought into the form
\begin{equation}
  \braket{B^0 |\Op_{\rm 6\,\,quarks}|\bq} \sim \braket{B^0
  |(\overline{h}_vd)(\overline{h}_vd) 
  (\overline{u}u) |\bq} \biggr[ \lambda_t^2 - \lambda_c
  \lambda_t \left( \frac{m_c^2}{m_b^2} + \xi \right) + \Op
  (\frac{m_c^4}{m_b^4}) \biggr],   
\end{equation}
where $\xi$ is defined as 
\begin{equation}
  \frac{\braket{B^0|(\overline{h}_vd)(\overline{h}_vd) (\overline{u}u)
  |\bq}}{\braket{B^0|(\overline{h}_vd)(\overline{h}_vd) 
  (\overline{c}c) |\bq}} = 1 + \xi . 
\end{equation}
The GIM mechanism
as well as the OPE guarantee, that $\xi$ is of the same order as 
the $m_c^2/m_b^2$ terms. Altogether, the dim-9 operators are
suppressed with respect to the dim-6 operators by a factor of
$\Lambda^2_{\text{QCD}} / m_b^2$, as could have been guessed from the
OPE. Hence they are a negligible contribution, at most of the absolute
order of $10^{-6}$. 

Another class of corrections are the $\mathcal{O}(\alpha_s)$ QCD
corrections which are  
known for most of the processes at the two-loop level
\cite{weakdecays}. For the box diagrams  
they have been already calculated some time ago 
(see the references in \cite{weakdecays}). However, 
one may use the effective field theory to resum large logarithms of the form 
$\alpha_s \ln (M_W/ \mu)$, where $\mu$ is a hadronic scale. 
Since there are large logarithms already at order 
$(\alpha_s)^0$ the situation is similar to the one in the transition $s \to d \ell \ell$,
which has been discussed in \cite{GilmanWise}. 

The result given in (\ref{simple}) does not resum the logarithms of the form 
$\alpha_s \ln (M_W/ \mu)$. For the case of the $\Delta S = 2$ 
effective Hamiltonian, the next-to-leading result has been given in 
\cite{NiersteHerrlich}, however, in our case the situation is slightly
different 
due to the fact that the mass of the bottom quark sets a large scale and thus 
a matching to an effective theory with a static $b$ quark at the scale 
$\mu \sim m_b$ is possible, which is still  perturbative.  The running below
the scale $m_b$ down to the scale $m_c$ resums logarithms of the form 
$\ln m_b^2 / m_c^2$; however, these logarithms are not large and hence
the resummation is not really needed. In principle one could run perturbatively 
even  below the charm mass, yielding a result like (\ref{simple}) in terms of local 
operators, but the running below $m_b$ is a very small effect which we shall 
neglect in the following. 

Working at one loop requires to include the one-loop correction to the 
$\Delta B = 1$ effective Hamiltonian as well as the mixing among the local 
$\Delta B = 2$ operators of dimension 8. However, since we are only interested 
in the contributions that modify the relation between $\sin (2 \beta)$ and 
$S_{B \to J/\Psi K_s}$ we shall simplify the discussion by neglecting the 
mixing among the $\Delta B = 2$ operators of dimension 8, since only the 
single dim-8 operator proportional to $m_c^2$ will contribute to this effect. 

The running  of the ${\Delta B = 1}$ effective Hamiltonian forces us to 
introduce operators $T_i$ with different color combinations. It is well known 
that for the $\Delta B = 1$ operators the renormalization-group evolution is 
diagonalized by the combinations \cite{GilmanWiseOld}
\begin{equation}
  \left( (\bar{b}_L \gamma_\mu q_L ) (\bar{q}^{\,\prime}_L
  \gamma^\mu d_L ) \right)_\pm  = \frac12 \left[  (\bar{b}_{i,L}
  \gamma_\mu q_{i,L} ) 
  (\bar{q}^{\,\prime}_{j,L} \gamma^\mu d_{j,L} ) \pm 
  (\bar{b}_{i,L} \gamma_\mu q_{j,L} )
  (\bar{q}^{\,\prime}_{j,L} \gamma^\mu d_{i,L} ) \right] . 
\end{equation}
Thus it is convenient to introduce the non-local operators in the form 
\begin{equation}
T^{\sigma\sigma'}_{qq'} =  -\frac{i}{2} \int d^4 x \,
T \biggl[ \left( (\bar{b}_L \gamma_\mu q_L )(x) (\bar{q}^{\,\prime}_L
  \gamma^\mu d_L )(x) \right)_\sigma 
  \left( (\bar{b}_L \gamma^\nu q'_L )(0) (\bar{q}_L \gamma_\nu d_L
  )(0) \right)_{\sigma'} \biggr] ,
\end{equation}
where  $\sigma$ and $\sigma'$ may take the values $+, -$.
Ignoring the mixing among the local dim-8 operators we choose as the basis
\begin{equation*}
\vec{O} =  (Q_3, T^{++}_{cc}, T^{+-}_{cc},
T^{--}_{cc}, T^{++}_{cu}, T^{+-}_{cu}, T^{--}_{cu}, T^{++}_{uu},
T^{+-}_{uu}, T^{--}_{uu})^T.
\end{equation*}

For the case at hand the relevant contribution is the mixing of the operators 
from the time-ordered products into the local operator $Q_3$, for which we also 
keep the mixing with itself. The anomalous dimension matrix, including tree level 
and this restricted set of  $\alpha_s$ corrections,  becomes
\begin{eqnarray} \label{anodim}
\gamma =  
      \setlength{\extrarowheight}{2pt}
      \setlength{\arraycolsep}{2pt}
    \left(\begin{array}{c||ccc|ccc|ccc}
          -\ap & 0 & 0 & 0 & 0 & 0 & 0 & 0 & 0 & 0 \\\hline\hline
	   \frac{3}{16\pi^2}  & 2\ap & 0 & 0 & 0 & 0 & 0 & 0 & 0 & 0 \\ 
          -\frac{1}{16\pi^2} & 0 & -\ap & 0 & 0 & 0 & 0 & 0 & 0 & 0 \\
           \frac{1}{16\pi^2} & 0 & 0 & -4\ap & 0 & 0 & 0 & 0 & 0 & 0 \\ \hline
	   \frac{3}{32\pi^2} & 0 & 0 & 0 & 2\ap & 0 & 0 & 0 & 0 & 0 \\ 
	  -\frac{1}{32\pi^2}  & 0 & 0 & 0 & 0 & -\ap & 0 & 0 & 0 & 0 \\ 
           \frac{1}{32\pi^2} & 0 & 0 & 0 & 0 & 0 & -4\ap & 0 & 0 & 0 \\\hline 
           0  & 0 & 0 & 0 & 0 & 0 & 0 & 2\ap & 0 & 0 \\
	   0 & 0 & 0 & 0 & 0 & 0 & 0 & 0 & -\ap & 0 \\
           0 & 0 & 0 & 0 & 0 & 0 & 0 & 0 & 0 & -4\ap \\
        \end{array}\right).
\end{eqnarray}

In terms of the ``diagonalized''  coefficients
\begin{equation}
\begin{split}
  C^{++} &= \; C_{22} + C_{11} + C_{12} + C_{21}  \\
  C^{--} &= \; C_{22} + C_{11} - C_{12} - C_{21}  \\
  C^{+-} &= \; C_{22} - C_{11} + C_{12} - C_{21}  \\
  C^{-+} &= \; C_{22} - C_{11} - C_{12} + C_{21} 
\end{split}
\end{equation}
the matching conditions (i.e.\  the starting points for the
renormalization-group evolution) are 
\begin{align}
  C_{cc}^{++} (M_W^2)&= \lambda_c^2 , & C_{cc}^{+-} (M_W^2)&= 
  2 \lambda_c^2, & C_{cc}^{--} (M_W^2)&=  \lambda_c^2, \notag \\ 
  C_{cu}^{++} (M_W^2) &= 2 \lambda_c\lambda_u , & C_{cu}^{+-} (M_W^2) &= 
  4 \lambda_c\lambda_u, & C_{cu}^{--} (M_W^2) &= 2 \lambda_c\lambda_u,
   \\ 
  C_{uu}^{++} (M_W^2) &=  \lambda_u^2, & C_{uu}^{+-} (M_W^2) &= 
  2 \lambda_u^2, & C_{uu}^{--} (M_W^2) &=  \lambda_u^2, \notag
\end{align}
where we combined $+-$ and $-+$. 

The solution of the renormalization-group equation (at one loop) 
for the Wilson coefficients of the $\Delta B = 1$ operators is well
known
($\gamma^{\sigma\sigma'} =
\tg^{\sigma\sigma'} \alpha_s(\mu)/\pi$,
$\tg^{++,+-,--} = 2,-1,-4$) and yields 
\begin{equation}
  C^{\sigma\sigma'}_{qq'}(\mu) = C^{\sigma\sigma'}_{qq'} (M_W) \cdot
  \left( \dfrac{\alpha_s(\mu)}{\alpha_s(M_W)} \right)^{-2
  \tg^{\sigma\sigma'}/\beta_0},
\end{equation} 
where $\beta_0 = 11 - 2 N_f/3$ is as usual the 
coefficient of the one-loop QCD beta function. Due to the special structure of the
anomalous dimension matrix, we obtain the renormalization-group
equation for the coefficient $C_3$ of the operator $Q_3$   
\begin{align}
  \mu \frac{d}{d\mu} C_3(\mu) &= - \frac{\alpha_s(\mu)}{\pi} C_3(\mu)
  - \frac{\lambda_c\lambda_t}{16 \pi^2}  \sum_{\xi=++,+-,--}
  \Gamma^\xi \left(
  \dfrac{\alpha_s(\mu)}{\alpha_s(M_W)} \right)^{-2
  \tg^\xi/\beta_0} ,
\end{align}
where  $\Gamma^{++,+-,--} = 3,-2,1$ originates from the first column of 
(\ref{anodim}). 
Note that there is a factor two for $+-$ due to the equal contribution
from $+-$ and $-+$. Furthermore,  the flavor dependence of the
time-ordered products' mixing into the quasi-local operators appeared
within the structure $\lambda_c^2 + \frac12 \cdot 2 \lambda_u\lambda_c
= -\lambda_c \lambda_t$. The factor of two here stems from the fact that
the $uc$ and $cu$ flavor combinations give the same contribution. 

The equation for $C_3$ can be solved by standard methods;  
the solution of the homogeneous differential equation is 
\begin{equation}
  C^{(0)}_3(\mu) = c \cdot
  \left( \dfrac{\alpha_s(\mu)}{\alpha_s(M_W)} \right)^{-2
  \tg_3/\beta_0} .  
\end{equation}
By setting $C_3(\mu) =
c(\mu)C_3^{(0)}(\mu)$ we get the solution 
\begin{multline}
  C_3(\mu) = \left( \dfrac{\alpha_s(\mu)}{\alpha_s(M_W)} \right)^{-2
  \tg_3/\beta_0} \cdot \Biggl\{ \hat{C}_3(M_W,x_t) + \frac{\lambda_c
  \lambda_t}{8 \pi \beta_0} \sum_\xi \Gamma^\xi \left(
  \frac{2(\tg_3 - \tg^\xi)}{\beta_0} - 1 \right)^{-1}
   \cdot \\ \qquad  \biggl[ \frac{1}{\alpha_s(\mu)}
  \left(\frac{\alpha_s(\mu)}{\alpha_s(M_W)}\right)^{2 \frac{(\tg_3 -
  \tg^\xi)}{\beta_0}}  - 
  \frac{1}{\alpha_s(M_W)}\biggr] \Biggr\} .
\end{multline}

We shall consider only the evolution from $M_W$ down to $m_b$, since at $m_b$ 
we would need to consider again a different set of operators,
including static quarks for the $b$ and later also for the $c$  and their
renormalization \cite{GrinsteinKilian}. While this can be done in principle, 
the corresponding logarithms $\alpha_s \ln(m_b/m_c)$ and $\alpha_s \ln(m_c/\mu)$ 
are smaller than the ones from the running from $M_W$ to $m_b$ and we shall 
include here only the leading term. 

We obtain for the QCD corrections at $\mu \sim m_b$ the explicit formula
\begin{multline}
  \Delta \Im\left[\frac{M_{12}}{|M_{12}|} \right] = \frac{1}{C_0(x_t)}
  \frac{m_c^2}{m_t^2} \frac{|V_{cd}|
  |V_{cb}|}{|V_{td}| |V_{tb}|} \left(
  \dfrac{\alpha_s(\mu)}{\alpha_s(M_W)} \right)^{-2\tg_3/\beta_0} 
  \Biggl\{ \frac{|V_{cd}|
  |V_{cb}|}{|V_{td}| |V_{tb}|} \sin(2\beta)\cos(2\beta) 
  \\ + 2\biggl( C_3(x_t) + \frac{2 \pi}{\beta_0} \sum_\xi
  \Gamma^\xi \left( 
  \frac{2(\tg_3 - \tg^\xi)}{\beta_0} - 1 \right)^{-1}
   \cdot \biggl[ \frac{1}{\alpha_s(\mu)}
  \left(\frac{\alpha_s(\mu)}{\alpha_s(M_W)}\right)^{2 \frac{(\tg_3 -
  \tg^\xi)}{\beta_0}}  - 
  \frac{1}{\alpha_s(M_W)}\biggr] \biggr) \\
   \cdot 
  \left(\sin\beta - \sin(2\beta) \cos\beta\ \right)
   \Biggr\}  ,
\end{multline}
which turns into the simple case  (cf.~(\ref{simple2})) as $\alpha_s (M_W) \to 0$. 
Numerically, one obtains at $\mu = m_b$ 
\begin{equation}
  \Delta \Im\left[\frac{M_{12}}{|M_{12}|}
    \right]_{\rm QCD}   = - 2.08 \cdot 10^{-4},  
\end{equation} 
which has to be compared to 
\begin{equation}
  \Delta \Im\left[\frac{M_{12}}{|M_{12}|} \right] =  - 3.00 \cdot 10^{-4}
\end{equation}
without QCD corrections. This indicates that QCD can reduce the
absolute value of the correction to $\sin (2 \beta)$ coming from mixing
by about 30\%. Although we have not included the mixing among the 
dim-8 operators, we still take this as the size of the QCD corrections to 
be expected. We conclude that these contributions are safely  
below the percent level. 


\section{Corrections to the decay}
The operators contributing to the decay  $B^0$ to $J/\Psi K_S$ are
all of the flavor structure $(\bar{b} q)(\bar{q} s)$ where at
leading order we have from the current-current operators  
$q = c,u$. From these two operators $q=c$ is Cabibbo-favored
over $q=u$. Furthermore, also the penguin contributions are dominated by the
charm quark, such that all contributions to the decay carry the same
weak phase and hence one expects a very small direct CP asymmetry \cite{BigiSanda}.
Note that the top quark contribution has been integrated out already at the 
weak scale for both the QCD and the electroweak penguins.

However, looking for small deviations we have to study the small terms
carrying different weak phases which is the $(\bar{b} u)(\bar{u} s)$
contribution. Aside from the QCD
penguins also electroweak penguins become important once small corrections
are studied \cite{RFewp}. We identify the corresponding matrix elements of the
two different contributions -- tree and penguin -- to the 
effective Hamiltonian in the numerator and denominator of the ratio
of amplitudes  
\begin{equation}
  \frac{A(\bq \to J/\Psi K_S)}{A(B^0 \to J/\Psi
  K_S)} = \frac{\braket{J/\Psi K_S|\mathcal{T}_{\text{eff}} (b \to c\bar{c}s)
  |\bq} + \braket{J/\Psi K_S|\mathcal{T}_{\text{eff}} (b
  \to u\bar{u}s) |\bq}|\xi_u/\xi_c| e^{-\ii\gamma}}{\braket{J/\Psi
  K_S|\mathcal{T}_{\text{eff}} (b \to c\bar{c}s) 
  |B^0} + \braket{J/\Psi K_S|\mathcal{T}_{\text{eff}} (b \to u\bar{u}s)
  |B^0}|\xi_u/\xi_c|e^{+\ii\gamma}},
\end{equation}
where $\mathcal{T}_{\text{eff}}  (b \to q\bar{q}s) $ is the sum over the operators 
(multiplied by their Wilson Coefficients) with the quark content $(b \to q\bar{q}s)$. 
It is convenient to define the ratio  
\begin{equation}
  r := \frac{\braket{J/\Psi K_S|\mathcal{T}_{\text{eff}} (b \to u\bar{u}s)
  |\bq}}{\braket{J/\Psi
  K_S|\mathcal{T}_{\text{eff}} (b \to c\bar{c}s)|\bq}}
  \left|\frac{\xi_u}{\xi_c}\right|, \qquad
  \left|\frac{\xi_u}{\xi_c}\right| = \frac{|V_{ub}| |V_{us}|}{|V_{cb}|
  |V_{cs}|} = 0.0203 \pm 0.0066
\end{equation}
as an expansion parameter. 
In this way we get
\begin{equation}
  \eta_{J/\Psi K_S} \cdot \frac{A(\bq \to J/\Psi K_S)}{A(B^0 \to J/\Psi
  K_S)} = \frac{1 + r e^{-\ii\gamma}}{1 + r
  e^{+\ii\gamma}} \approx 1 - 2 \, \ii \, r \sin\gamma,
\end{equation}
where $\eta_{J/\Psi K_S} = -1$ is the CP eigenvalue. 

The main obstacle to obtain a reliable quantitative estimate is the evaluation of
these hadronic matrix elements. Some time ago the 
so-called BSS mechanism~\cite{BSS} has been suggested, 
where the $\overline{u} u$-loops
are evaluated perturbatively, assuming a sufficiently large momentum
transfer through this loop. In fact, this approach has been supported
recently by QCD factorization \cite{BBNS}, indicating that the loop is indeed 
perturbative.
\vspace{5mm}
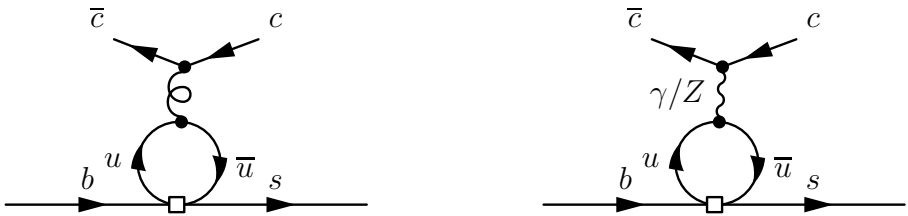
\begin{equation*}
  \parbox{55\unitlength}{\hfil\\%
      \begin{fmfgraph*}(48,22)
        \fmfstraight
        \fmfleft{l}
        \fmfright{r}
        \fmftopn{t}{11}
        \fmfbottomn{b}{20}
        \fmf{phantom}{b10,dum1,dum2,dum3,dum4,dum5,dum6,dum7,t6} 
        \fmf{fermion,left,tension=0.4,label=$u$,l.side=left}{b10,dum5}
        \fmf{fermion,left,tension=0.4,label=$\overline{u}$,l.side=left}{dum5,b10}
        \fmf{fermion,label=$b$,l.side=left}{b1,b10}
        \fmf{fermion,label=$s$,l.side=left}{b10,b20}
        \fmf{fermion}{t8,v1,t4}
        \fmf{gluon}{v1,dum5}
        \fmfdot{dum5,v1}
        \fmfv{decor.shape=square,decor.filled=empty,decor.size=2mm}{b10}
        \fmflabel{$\overline{c}$}{t4}
        \fmflabel{$c$}{t8}
      \end{fmfgraph*}\hfil\\} \qquad\qquad
  \parbox{55\unitlength}{\hfil\\%
      \begin{fmfgraph*}(48,22)
        \fmfstraight
        \fmfleft{l}
        \fmfright{r}
        \fmftopn{t}{11}
        \fmfbottomn{b}{20}
        \fmf{phantom}{b10,dum1,dum2,dum3,dum4,dum5,dum6,dum7,t6} 
        \fmf{fermion,left,tension=0.4,label=$u$,l.side=left}{b10,dum5}
        \fmf{fermion,left,tension=0.4,label=$\overline{u}$,l.side=left}{dum5,b10}
        \fmf{fermion,label=$b$,l.side=left}{b1,b10}
        \fmf{fermion,label=$s$,l.side=left}{b10,b20}
        \fmf{fermion}{t8,v1,t4}
        \fmf{photon,label=$\gamma/Z$}{v1,dum5}
        \fmfdot{dum5,v1}
        \fmfv{decor.shape=square,decor.filled=empty,decor.size=2mm}{b10}
        \fmflabel{$\overline{c}$}{t4}
        \fmflabel{$c$}{t8}
      \end{fmfgraph*}\\\hfil} 
\end{equation*}

The evaluation of the loop requires to insert a 
typical momentum
transfer $k^2$ passing through the up-quark loop \cite{BSS}; furthermore, the loop also
depends on the typical scale $\mu$ of the problem, which will be the
mass of the $b$ quark. Since the loop involves only scales well
above the hadronic scale, one obtains again local operators and thus
one may use as an estimate   
\begin{multline}
  \mathcal{H}^{\text{Peng.}}_{\text{eff}}(b\to c\bar{c}s)= -
  \frac{G_F}{\sqrt{2}} \Biggl\{
  \frac{\alpha}{3\pi} \left( 
  \overline{s} b \right)_{V-A} \left( \overline{c} c \right)_V \cdot 
  \left[ 1 + \mathcal{O}\left(\frac{M_\Psi^2}{M_Z^2}\right) 
  \right]   \\ +
  \frac{\alpha_s}{3\pi} \left( \overline{s} T^a b \right)_{V-A} \left(
  \overline{c} T^a c \right)_V \Biggr\} \cdot  
  \left( \frac{5}{3} - \ln \left(
  \frac{k^2}{\mu^2} \right) + \ii \pi \right),
\end{multline}
where the first term originates from the
electroweak penguin, the second one from the QCD penguin.

The remaining problem is to estimate the color-singlet and the
color-octet matrix element. We shall take a simple-minded approach
and estimate the sizes of these matrix elements from the total rate of
the decay $B^0 \to J/\Psi K_S$. From the measured  
$B^0$ life time, $\tau = (1.537 \pm 0.015) \,\text{ps}$ \cite{PDG},
and the branching ratio $\text{BR}(B^0 \to J/\Psi K_S) = (4.25 \pm
0.25)\cdot 10^{-4}$ \cite{PDG} one can calculate the matrix element for the
$\overline{c}c$-contribution by taking the square root
\begin{equation}
  |\braket{J/\Psi
   K_S|(\overline{s}b)(\overline{c}c)|\overline{B}^0}|_{\text{exp.}} =
   (8.36 \pm 0.66) \cdot 10^8 \; \text{MeV}^3 .  
\end{equation}

The matrix element with the charm pair being in the octet is estimated
by splitting the effective Hamiltonian in singlet and octet
contributions. In naive factorization only the singlet piece survives
which, however, yields a rate roughly a factor three to four too low
compared to experiment. The difference comes from nonfactorizable
contributions which we shall ascribe completely to the octet term. 
To obtain an estimate of this matrix element from the measured total
rate we also need the relative phase of the two contributions, which we
take from QCD light-cone sum rule estimates in~\cite{melic} (see
also~\cite{sumrules}). The relative phase turns out to be small and
so we may simply add the two parts. 

For the singlet matrix element calculated by means of factorization we get 
\begin{equation}
  |\braket{J/\Psi
   K_S|(\overline{s}b)(\overline{c}c)|\overline{B}^0}|_{\text{fact.}}
   = (3.96 \pm 0.36) \cdot 10^9 \; \text{MeV}^3 .   
\end{equation}
Inserting for the Wilson coefficients the values\footnote{We take the Wilson
    coefficients at Leading-Log order at the scale $m_b$; unfortunately,
    the scale dependence is still large at leading order, but we are aiming
    only at an estimate.} 
\begin{equation}
  C^{(1)} = C_1 + \frac{1}{N_c} C_2 = 0.10 \pm 0.03 , 
  \qquad\quad C^{(8)} = 2 C_2 = 2.24 \pm 0.04 ,
\end{equation}
we get as an estimate for the octet matrix element
\begin{equation}
 |\braket{J/\Psi
   K_S|(\overline{s}T^a b)(\overline{c}T^a c)|\overline{B}^0}| = (1.97
   \, \pm \, 0.64) \cdot 10^8 \, \text{MeV}^3 .
\end{equation}

Taking the scale to be $m_b$ and momentum transfer as
$m_{J/\Psi}$ gives (the two contributions are the electroweak and the
QCD penguins, respectively) 
\begin{multline}
  r = \left[-(0.16 + 1.27) \, \pm 0.66\right] \cdot 10^{-4} \;\; \cdot
  \left( \frac{5}{3} -  \ln \left(
  \frac{k^2}{\mu^2} \right) + \ii \pi \right) \quad \Rightarrow \\ 
  \Re\left[ r \right] = \left( - 3.62 \, \pm 1.55 \right)
  \cdot 10^{-4} , \qquad 
  \Im\left[ r \right] = \left( - 4.48 \, \pm 1.92 \, \right)
  \cdot 10^{-4} .
\end{multline}

We have to point out that the estimates of these matrix elements are extremely 
difficult and hence quite uncertain; the electroweak penguins have been estimated 
in a recent paper with a similar approach \cite{BFRS}. 

\section{Effects on the CP asymmetry}

Now we are ready to collect all elements for the CP violation terms
within the asymmetry. The quantity  $|\lambda|^2$ is close to unity, 
the deviation being a small quantity of the same order as $r$. Thus we
write $|\lambda|^2 =: 1 + \delta$ and get 
\begin{align}
   C_{B\to J/\Psi\,K_S}
  \approx&\; - \frac{\delta}{2}.
\end{align}
Therein $\delta$ is given by
\begin{equation}
  \delta = \Delta A + \epsilon = 4 \, \Im\left[ r \right] \,
  \sin\gamma - \Im \left[ 
  \frac{\Gamma_{12}}{M_{12}} \right],
\end{equation}
where $\Delta A$ is the deviation of the $|\overline{A}/A|^2$ from 1
and $\epsilon$ has been defined in (\ref{defepsilon}). 

For the ratio between the off-diagonal width and mass matrix elements
one can take e.g. the calculation in \cite{Bigi92}\footnote{Here we
are assuming the solution for $\beta$ in accord with the unitarity triangle.}:
\begin{equation}
  \epsilon = - \Im \left[ \frac{\Gamma_{12}}{M_{12}} \right] = -
  \frac{4 \pi}{C_0 (x_t)} \frac{m_c^2}{m_t^2} \, \Im \left[ \frac{V_{cb}
  V_{cd}^*}{V_{tb} V_{td}^*} \right] \approx + \left( 5.18 \pm 2.96
  \right) \cdot 10^{-4} , 
\end{equation}
from which we obtain a numerical value for  $\delta$ 
\begin{equation}
  \delta = - \left( 1.02 \, \pm \, 0.75 \right) \cdot 10^{-3} .\;\;\;
\end{equation}

Including all small corrections, the  time-dependent CP asymmetry in 
$B^0 \to J/\Psi K_S$ has to be fitted to 
\begin{equation} \label{fit}
\begin{split}
  a_{\text{CP}} (t) &=\;
  - \left( \sin(2\beta) + \Delta S_{B\to
  J/\Psi\,K_S} \right) \cdot \sin(\Delta m \, t)  -
  \frac{\delta}{2} \cos(\Delta m \, t) \\ & \qquad\quad
  +\frac{\epsilon}2 - \frac{\epsilon}2  
  \sin^2(\Delta m \, t)\sin^2(2\beta)
  + \sin(4\beta) 
  \frac{\Delta\Gamma \, t}{4}  \sin(\Delta m \, t).
\end{split}
\end{equation}
Besides the correction to the leading $\sin(2\beta)$ term and the
cosine term there is now a small constant contribution to the CP
asymmetry as well as terms proportional to $\sin^2(\Delta m \,t)$ and 
$t \sin(\Delta m\,t)$. 

The correction to the mixing-induced CP violation is (note that the
term containing the imaginary part of $r$ cancels in the expression for the
correction)  
\begin{equation}
\begin{split}
  \Delta S_{B\to J/\Psi\,K_S} 
  &=\; 2 \, \Im\left[ r \right] \, \sin\gamma \, \sin(2\beta) -
  \Delta \Im\left[ 
  \frac{M_{12}}{|M_{12}|} \right] + 2 \sin\gamma \, \Re\left[ r 
  \frac{M^*_{12}}{|M_{12}|} \right] \\ &=\; 2 \sin\gamma \, \Re\left[ r \right]
  \cos(2 \beta) - \Delta \Im\left[
  \frac{M_{12}}{|M_{12}|} \right] .
\end{split}
\end{equation}

With the most recent value of $\sin(2\beta) = 0.736 \pm 0.49$
\cite{sin2beta}, the corrections in $\Delta S_{B\to J/\Psi\,K_S}$ have
the size of 
\begin{subequations}
\begin{align}
  \Delta \Im\left[\frac{M_{12}}{|M_{12}|} \right] &=\;  
  -(2.08 \pm 1.23) \cdot 10^{-4} \\ 
  2 \sin\gamma \, \Re\left[ r \right]
  \cos(2 \beta)  &=\;  - (4.24 \pm 1.94) \cdot 10^{-4}
\end{align} 
\end{subequations}
and sum up to 
\begin{equation}
   \Delta S_{B\to J/\Psi\,K_S} = - (2.16 \pm
   2.23) \cdot 10^{-4},
\end{equation}
which is a correction of roughly a third of a per mil with respect to the
measured value for $\sin(2\beta)$, but due to the large uncertainty could also 
be much smaller. 

From the mass difference as measured quantity 
and the approximate calculation for
\begin{equation}
  \label{eq:deltagam}
\Delta\Gamma \approx - \frac{3\pi}{2 C_0(x_t)} \frac{m_b^2}{m_t^2}
  \left[ 1 - \frac83 \frac{m_c^2}{m_b^2} \frac{|V_{cb}| |V_{cd}|}{|V_{tb}|
  |V_{td}|} \cos\beta \right] \Delta M_B,
\end{equation}
one can determine the term leading to a linear dependence 
on $t$ in (\ref{fit}) 
\begin{align}
  \Delta \Gamma &\approx - \left( 1.773 \pm 0.249 \right) \cdot 10^{-12}
  \;\text{MeV} \notag\\
  &\approx 
  - \left( 2.694 \pm 0.378 \right) \cdot 10^{-3}
  \;\text{ps}^{-1} .
\end{align}
Hence, this term has a typical size of the order of
\begin{equation}
  \sin(4\beta) 
  \frac{\Delta\Gamma \tau_{B^0}}{4} \approx -(1.03 \pm 0.15) \cdot
  10^{-3}  \;\;, 
\end{equation}
again a very small contribution, which, in addition, will not show up at 
small times since its time dependence is $(t/\tau_{B^0}) \sin(\Delta m \, t)$.  

\section{Discussion}
We have reinvestigated the well known fact, that the mixing-induced CP asymmetry
of $B^0  \to J/\Psi K_S$ provides us with a very clean measurement of
$\sin (2 \beta)$.  
Already in the original paper \cite{BigiSanda} it was argued that in the 
standard model the contamination 
from ''wrong'' weak phases is tiny in this decay, but in the meantime the 
measurements became so precise that we considered it worthwhile to 
attempt again to quantitatively analyze these small standard-model contributions. 
In particular, we have used an effective field theory ansatz for the analysis of 
$\Delta B = 2$, which has not been employed before to this process. 

Our motivation was twofold. First of all, 
it has been argued that the CP asymmetry could give a hint to new physics, which 
has been analyzed generically in \cite{FleischerMannel}. It turns out that the 
general picture conjectured  in \cite{FleischerMannel} is supported by the present 
analysis. Secondly, the $B$ factories 
are doing very well and produce a large amount of data. Already now the 
measurement  of the CP asymmetry in the gold-plated mode is a precision 
measurement with uncertainties at the level of less than ten percent. In the near 
future this measurement will improve further. 

We have shown that the corrections to be expected in the standard model can 
partially  be calculated systematically, namely the part originating
from corrections to 
the mixing. Unfortunately, the second contribution, which is the one
from the decay matrix
element, is much harder to access, so we still cannot obtain a reliable estimate 
for the corrections to the CP asymmetry of the gold-plated mode. This situation 
could be improved once data on the decay $B_s \to J/\Psi K_S$ becomes available, 
since some of the uncertainties could be eliminated using these data \cite{RFBs}. 
However,  currently one has to use the 
methods that have been proposed by different authors and one can   
infer that the corrections will be very small, in the range of a few
per mil.  At least  
we can conclude from our analysis that there is still room for new physics in this 
observable, since a deviation from the standard-model prediction at the level of 
percents (which is not yet the experimental accuracy) 
would indicate the presence of new physics. 


\subsection*{Acknowledgements}
This work has been supported by the DFG Sonderforschungsbereich (SFB)
``Transregio 9 -- Computergest\"utzte Theoretische Teilchenphysik''
and the Graduiertenkolleg (GK) ``Hoch\-energiephysik und
Teilchenastrophysik''. TM is also supported by the BMBF under contract
05 HT1VKB/1. We are grateful to R.~Fleischer, A.~Grozin,
A.~Khodjamirian, B.~Meli\'c, U.~Nierste and T.~Selz for valuable
discussions.  

\end{fmffile}

\end{document}